\documentclass[aps,prl,twocolumn,groupedaddress]{revtex4-1}
\usepackage{amsmath,amsfonts,mathtools,amssymb}
\usepackage[colorlinks=true,linkcolor=blue,citecolor=red,bookmarksnumbered=true]{hyperref}
\usepackage{times,mathpazo,charter,graphicx,bm,enumitem}
\usepackage{color}

\makeatletter
\def\overl@ss#1#2{\vcenter{\offinterlineskip
        \ialign{$\m@th#1\hfil##\hfil$\crcr#2\crcr<\crcr } }}
\def\overgr@at#1#2{\vcenter{\offinterlineskip
        \ialign{$\m@th#1\hfil##\hfil$\crcr#2\crcr>\crcr } }}
\def\gl{\mathrel{\mathpalette\overl@ss>}}
\def\lg{\mathrel{\mathpalette\overgr@at<}}
\makeatother

\def\re{{\mathrm{re}}}
\def\im{{\mathrm{im}}}

\def\sn{\mathop{\rm sn}\nolimits}
\let\^=\hat
\let\==\bar
\let\@=\mathbf

\def\e{\mathrm{e}}

\def\~#1{\tilde{\mathbf{#1}}}
\def\txtfrac#1#2{{\textstyle\frac{#1}{#2}}}

\let\Delta=\varDelta

\def\[{\begin{equation}}
\def\]{\end{equation}}
\def\be{\begin{equation}}
\def\ee{\end{equation}}
\def\bse{\begin{subequations}}
\def\ese{\end{subequations}}

\newcommand\partialderiv[3][]{\frac{\partial^{#1}#2}{\partial {#3}^{#1}}}

\long\def\paragraph#1{\par\medskip\textbf{#1.}}

\allowdisplaybreaks

\def\reftitle#1{``#1''}

\newdimen\figwdlt
\newdimen\figwdrt
\newdimen\figvadjust
\newdimen\fighadjust
\newdimen\vtadjust
\long\def\leftrightfigurepanel#1#2{
\kern-\smallskipamount
\centerline{\kern-0.4em\includegraphics[width=\figwdlt]{#1}\kern-1.6em%
\raise0.6ex\hbox{\includegraphics[width=\figwdrt,angle=90]{#2}}}
\kern-\smallskipamount
}

\begin{document}

\title{Riemann problems and dispersive shocks in self-focusing media}
\author{Gino Biondini}
\affiliation{Department of Physics, State University of New York at Buffalo, Buffalo, New York 14260, USA\\
    Department of Mathematics, State University of New York at Buffalo, Buffalo, New York 14260, USA
    }
\date{\today}

%%%%%%%%%%%%%%%%%%%%%%%%%%%%%%%%%%%%%%%%%%%%%%%%%%%%%%%%%%%%%%%%%%%%%%%%%%%%%%%%%%%%%%%%%%%%%
\begin{abstract}
The dynamical behavior resulting from an initial discontinuity in focusing media is investigated 
using a combination of numerical simulations and Whitham modulation theory for the focusing nonlinear Schr\"odinger equation.
Initial conditions with a jump in either or both the amplitude and the local wavenumber are considered. 
It is shown analytically and numerically that the space-time plane divides into expanding domains in which the solution is described 
by a slow modulation of genus-zero, genus-one or genus-two solutions, 
their precise arrangement depending on the specifics of the initial datum.
\\[0.4ex]
\end{abstract}
\maketitle

%%%%%%%%%%%%%%%%%%%%%%%%%%%%%%%%%%%%%%%%%%%%%%%%%%%%%%%%%%%%%%%%%%%%%%%%%%%%%%%%%%%%%%%%%%%%%
\paragraph{Introduction}
The study of Riemann problems ---
i.e., the evolution of a jump discontinuity between two uniform values of the initial datum --
is a well-established part of fluid dynamics, 
since understanding the response of a system to such kinds of inputs 
is a key step in characterizing its behavior \cite{CH1962,Lax1973}.
When nonlinearity and dissipation are the dominant physical effects, these problems can give rise to classical shocks \cite{Lax1973,CF1976}.
When dissipation is negligible compared to dispersion, however,
Riemann problems can give rise to dispersive shock waves (DSWs).
The study of dispersive hydrodynamics has attracted considerable attention in recent years \cite{EH2016}.
%A comprehensive understanding of Riemann problems and DSW formation in self-focusing media, however, is still lacking.
%This work is aimed at overcoming this problem.
Here we are interested in Riemann problems and DSW formation in self-focusing media.

A universal model for nonlinear focusing media is the one-dimensional cubic nonlinear Schr\"odinger (NLS) equation,
which describes modulations of weakly nonlinear dispersive wave trains in many different physical contexts,
such as deep water waves, optical fibers, plasmas
and attracting Bose-Einstein condensates~\cite{AS1981,IR2000,Agrawal2001,PS2003}. 
The NLS equation is a completely integrable system,
and it can be studied via the inverse scattering transform (IST)~\cite{AS1981,NMPZ1984,FT1987,APT2004}.
Several related studies using IST have appeared in recent years
\cite{KMM2003,BV2007,BKS2011,BM2017,JM2013,BM2016,BM2017,PRE94p060201R,EKT2016}.
On the other hand, a simpler approach is
Whitham modulation theory \cite{Whitham1976,EH2016}.
The latter does not require integrability, and therefore it is not restricted to completely integrable
systems such as the NLS equation, and it can be extended to more general self-focusing dynamical systems such as those studied in \cite{SIREV}.

Whitham theory has been used extensively to study Riemann problems 
in defocusing media \cite{PHYSD1995v87p186,OL20p2291,SJAM59p2162,BK2006,HA2007}.
A few works have also used Whitham equations in focusing media \cite{elgurevich,bikbaev,kamchatnov}, 
and a few attempts have been made to combine it with IST \cite{bikbaev,EKT2016,ET2016}, 
but it is fair to say that the corresponding theory is much less developed than for defocusing media.
The main reason for this situation is that, while the Whitham equations for the defocusing NLS equation are hyperbolic,
those for the focusing NLS equation are elliptic, 
and are therefore unsuitable to study initial value problems (IVPs) in general.
Nonetheless, here we demonstrate how, despite this difficulty, a relatively broad class of Riemann problems 
can still be effectively studied using Whitham modulation theory.
%and we characterize the resulting phenomena.

%%%%%%%%%%%%%%%%%%%%%%%%%%%%%%%%%%%%%%%%%%%%%%%%%%%%%%%%%%%%%%%%%%%%%%%%%%%%%%%%%%%%%%%%%%%%%
\paragraph{The NLS equation and its Whitham equations}
We begin by reviewing some background material, to set up the relevant framework.
We write the one-dimensional focusing NLS equation with small dispersion as
\[
i\epsilon q_t + \epsilon^2 q_{xx} + 2|q|^2q = 0\,,
\label{e:nls}
\]
where subscripts $x$ and $t$ denote partial differentiation
and $0<\epsilon\ll1$ is a small parameter that quantifies the relative strength
of dispersion compared to nonlinearity.
Recall that Eq.~\eqref{e:nls} possesses several invariances: 
phase rotations, spatial reflections, scaling and Galilean transformation.
Specifically, if $q(x,t)$ is any solution of Eq.~\eqref{e:nls}, so are: 
$\e^{i\alpha}q(x,t)$, 
$q(-x,t)$,
$aq(ax,a^2t)$
and 
$\smash{\e^{i(Vx - V^2t)}}q(x-Vt,t)$,
where all transformation parameters are real-valued.
All of these invariances will be useful below.

As is well known, the so-called Madelung transformation, namely
\vspace*{-2ex}
\[
q(x,t) = \sqrt{\smash{\rho(x,t)}\phantom{d\!\!}}\,\e^{iS(x,t)/\epsilon},
%\label{e:madelung}
\]
where the real-valued quantities $\rho(x,t)$ and $S(x,t)$ represent respectively the local intensity and local phase,
transforms Eq.~\eqref{e:nls} into the hydrodynamic-type system
\unskip
\bse
\label{e:hydrodynamic}
\begin{align}
&\partialderiv \rho {t}= \partialderiv {(\rho v)}x\,, 
\\
&\partialderiv {(\rho v)}{t}= \partialderiv { }x \bigg(\,
  \rho v^2 - \frac12\rho^2 
  - \frac14\epsilon^2\rho\partialderiv[2]{}x\ln\rho \,\bigg)\,.
\end{align}
\ese
where $S_x(x,t) = v(x,t)$ is the local wavenumber.
The truncation $\epsilon=0$ of Eqs.~\eqref{e:hydrodynamic} is the genus-0 NLS-Whitham system.
Recall that Eq.~\eqref{e:nls} admits the background solution $q(x,t) = q_o\,\smash{\e^{2iq_o^2t}}$, 
together with its generalizations via the above-mentioned invariances.
The system~\eqref{e:hydrodynamic} describes slow modulations of such a solution.

In addition to constant and soliton solutions,
the focusing NLS equation~\eqref{e:nls} also admits solutions describing periodic, traveling wave solutions
in the form \cite{kamchatnov}
\vspace*{-1ex}
\begin{multline}
|q(x,t)|^2 = (\alpha_\im + \gamma_\im)^2
\\
- 4\alpha_\im\gamma_\im\,\sn^2\!\big[C(x-Vt);m\big],
\label{e:ellipticsolution}
\end{multline}
where $\sn(\cdot)$ is a Jacobi elliptic function \cite{NIST}, 
the elliptic parameter $m$ and the constant $C$ and $V$ are
\vspace*{-1ex}
%\bse
\begin{gather}
m = 
{4\alpha_\im\gamma_\im}/{|\alpha - \gamma|^2}\,, %= \frac{4\alpha_\im\gamma_\im}{(\alpha_\re - \gamma_\re)^2 + (\alpha_\im - \gamma_\im)^2},
\qquad
C = |\alpha - \gamma|\,, %= \sqrt{(\alpha_\re - \gamma_\re)^2 + (\alpha_\im - \gamma_\im)^2},
\end{gather}
%\ese
with $\alpha = \alpha_\re + i\alpha_\im$, $\gamma = \gamma_\re + i\gamma_\im$
Note that $\alpha$, $\alpha^*$, $\gamma$ and $\gamma^*$ are 
%the branch points of the elliptic function and also 
the branch points of the spectrum of the scattering problem associated to Eq.~\eqref{e:nls}
\cite{ItsKotlyarov1976}.

A system of modulation equations for the above periodic solutions of the focusing NLS Eq.~\eqref{e:nls} 

when $0<\epsilon\ll1$
can be obtained via
Whitham averaging theory \cite{Whitham1976}.
The result is the genus-1 NLS-Whitham system \cite{forestlee,pavlov}.
Explicitly, in Riemann invariant coordinates,
\vspace*{-\vtadjust}
\begin{gather}
\partialderiv{r_j}t + V_j\,\partialderiv{r_j}x = 0\,,\quad j = 1,\dots,4,
\label{e:genus1Whitham}
\end{gather}
where $r_1,\dots,r_4$ are the Riemann invariants, 
and the characteristic velocities are
\vspace*{-1ex}
\[
V_j = V + W_j\,,\quad j = 1,\dots,4\,,
\]
where $V = r_1 + r_2 + r_3 + r_4$, with
\bse
\begin{gather}
W_1 = 2\Delta_{12}/[1-(\Delta_{24}/\Delta_{14})\,R(m)]\,,
\\
W_3 = 2\Delta_{34}/[1-(\Delta_{24}/\Delta_{23})\,R(m)]\,,
\\
\Delta_{jk} = r_j - r_k\,,\quad j,k = 1,\dots,4\,,
\end{gather}
\ese
$W_2 = W_1^*$,
$W_4 = W_3^*$
and 
$R(m) = E(m)/K(m)$,
where $K(m)$ and $E(m)$ are the complete elliptic integrals of the first and second kind, respectively \cite{NIST}.
Importantly, the Riemann invariants are exactly the branch points of the elliptic solution~\eqref{e:ellipticsolution}.
That is, 
$r_1=\alpha$, $r_2=\alpha^*$, $r_3=\gamma$ and $r_4 = \gamma^*$.
\iffalse
Also, the average $\=\rho$ of $\rho(x,t)$ over one oscillation period is
\vspace*{-1ex}
\[
\=\rho = \rho_3 - (\rho_3-\rho_1)\,R(m)\,,
\]
with
\vspace*{-1ex}
\bse
\begin{gather}
\rho_1 = -\txtfrac14(r_1-r_2+r_3-r_4)^2 = -(\alpha_\re - \gamma_\re)^2\,,
\\
\rho_2 = -\txtfrac14(r_1-r_2-r_3+r_4)^2 = -(\alpha_\im - \gamma_\im)^2\,,
\\
\rho_3 = -\txtfrac14(r_1+r_2-r_3-r_4)^2 = (\alpha_\im + \gamma_\im)^2\,,
\end{gather}
\ese
with $m = \Delta_{12}\Delta_{34}/(\Delta_{13}\Delta_{24}) = (\rho_3 - \rho_2)/(\rho_3 -\rho_1)$.
\fi

The NLS Eq.~\eqref{e:nls} also admits multiphase solutions \cite{ItsKotlyarov1976}.  
In general, genus-$g$ solutions are expressed as ratios of Jacobi theta functions \cite{Zhou},
and their modulations are described by corresponding genus-$g$ Whitham modulation systems \cite{forestlee,pavlov}.
We refer the reader to \cite{EKT2016} for a review.  We will not use such higher-genus solutions and the corresponding modulation systems here.

%%%%%%%%%%%%%%%%%%%%%%%%%%%%%%%%%%%%%%%%%%%%%%%%%%%%%%%%%%%%%%%%%%%%%%%%%%%%%%%%%%%%%%%%%%%%%
\paragraph{Riemann problems for the focusing NLS equation}
The Whitham equations for Eq.~\eqref{e:nls} are elliptic,
and the Riemann invariants and the characteristic velocities are in general complex-valued.
Hence, these systems cannot be used to study IVPs in general,
contrary to the defocusing case.
Nonetheless, we next show that, notwithstanding this difficulty,
the system~\eqref{e:genus1Whitham} still yields useful information about the behavior of solutions of Eq.~\eqref{e:nls}.

We consider the focusing NLS equation with the following class of initial conditions (IC):
\vspace*{-1ex}
\[
q(x,0) = \begin{cases} 
    A_-\,\e^{i\mu x/\epsilon - i\phi/\epsilon}, &x<0,\\
    A_+\,\e^{-i\mu x/\epsilon + i\phi/\epsilon}, &x>0\,,
\end{cases}
\label{e:IC}
\]
with $A_\pm\ge0$ and $\mu$ and $\phi$ real,
and we classify the resulting dynamics depending on the values of $A_\pm$ and $\mu$.
(It turns out that the value of $\phi$ has no effect on our results.)
The results can be considered the analogue for the focusing case of those obtained in \cite{SJAM59p2162} for the defocusing case.

Nonzero values of $\mu$ correspond to the presence of carriers with opposite wavenumbers for $x<0$ and $x>0$, 
which, due to the Galilean invariance of the NLS equation, induce counter-propagating flows.
%The value of $\mu$ quantifies their strength. 
For $\mu>0$, the two halves of the IC~\eqref{e:IC} propagate inward (i.e., towards each other), 
whereas for $\mu<0$ they flow outward (i.e., away from each other).
Note that one can always take the discontinuity at $x=0$ 
and set the phases of the IC for $x<0$ and $x>0$ to be equal and opposite without loss of generality,
thanks to the translation and phase invariance of the NLS equation.
Similarly,
one can always take the carrier wavenumbers for $x<0$ and $x>0$ to be equal and opposite 
thanks to the Galilean invariance of the NLS equation.

%%%%%%%%%%%%%%%%%%%%%%%%%%%%%%%%%%%%%%%%%%%%%%%%%%%%%%%%%%%%%%%%%%%%%%%%%%%%%%%%%%%%%%%%%%%%%
\paragraph{One-sided step}
The simplest scenario of IC given by Eq.~\eqref{e:IC} is that of a one-sided step, in which 
$A_- = 0$. 
Then we can always set $A_+ >0$ and $\mu = \phi = 0$ without loss of generality 
thanks to the phase and Galilean invariances of Eq.~\eqref{e:nls}.
The long-time asymptotics of solutions generated by these IC was studied in \cite{BKS2011} by IST.
On the other hand, the dynamics can be effectively described via the genus-1 Whitham system~\eqref{e:genus1Whitham} \cite{EKT2016}.

Even though the Riemann invariants and the characteristic velocities are in general complex,
the genus-1 system~\eqref{e:genus1Whitham} does possess some real-valued solutions.
In particular, it admits the self-similar solution \cite{elgurevich}
\vspace*{-0.4ex}
\bse
\label{e:modulationsystem}
\begin{gather}
4\alpha_\re + 2(A_+^2-\alpha_\im^2)/{\alpha_\re} = \xi\,,
\qquad
\gamma = iA_+\,,
\\[0.4ex]
\big(\alpha_\re^2 + (A_+-\alpha_\im)^2\big)K(m) = (\alpha_\re^2-\alpha_\im^2+A_+^2)E(m)\,,
\label{e:modulation2}
\end{gather}
\ese
with $\xi = x/t$.\,\ 
This solution describes a slow modulation of the elliptic solution~\eqref{e:ellipticsolution} 
[now describing oscillations with characteristic spatial period $O(\epsilon)$,
as can be easily seen by performing a simple rescaling of the spatial and temporal variables in Eq.~\eqref{e:nls}
when $\epsilon\ne1$].
As discussed in \cite{EKT2016}, Eqs.~\eqref{e:modulationsystem} correctly capture the behavior of the solution of the NLS Eq.~\eqref{e:nls} 
with IC~\eqref{e:IC}
for $0<x<4\sqrt2 A_+ t$.
For $x>4\sqrt2 A_+ t$, the solution of the Whitham system is simply given by 
$\gamma = iA_+$ and $\alpha = 0$,
which matches the limit of the self-similar solution~\eqref{e:modulationsystem},
and which yields the constant solution $q(x,t) = A_+$
of the NLS equation (up to a uniform phase).
Finally, for $x<0$ the solution of the NLS equation is described by the degenerate, constant genus-0 solution 
$\gamma = \alpha =0$
of the 
Whitham system~\eqref{e:genus1Whitham},
which yields the trivial solution $q(x,t) = 0$ of the NLS equation.
Summarizing, the solution~\eqref{e:modulationsystem} describes an oscillatory wedge $V_-t < x < V_+t$,
with $V_-=0$ and $V_+ = 4\sqrt2 A_+$,
which connects the constant solution
$q(x,t) = 0$ to its left to the constant solution $q(x,t) = A_+$ to its right.

\begin{figure}[b!]
\leftrightfigurepanel{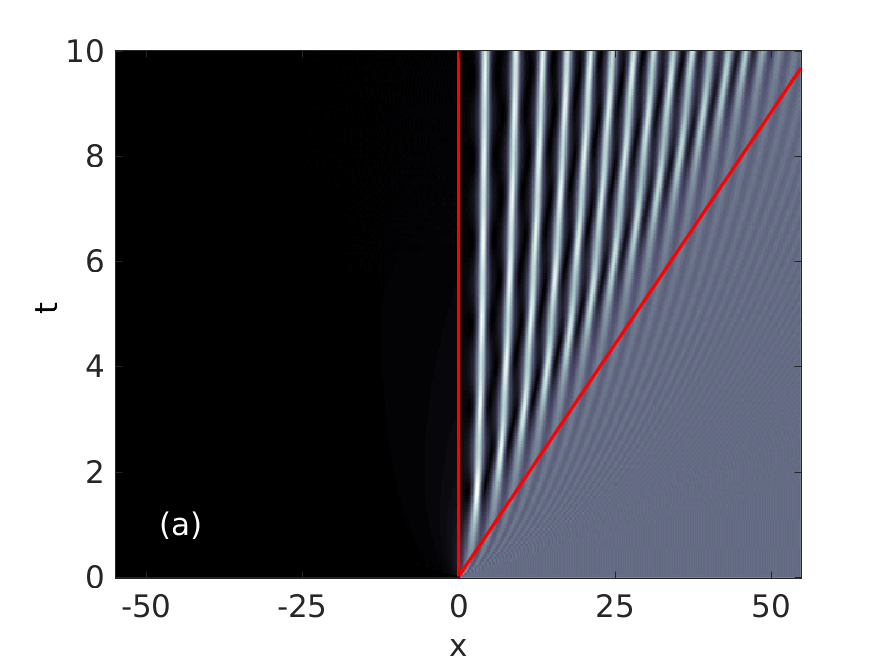}{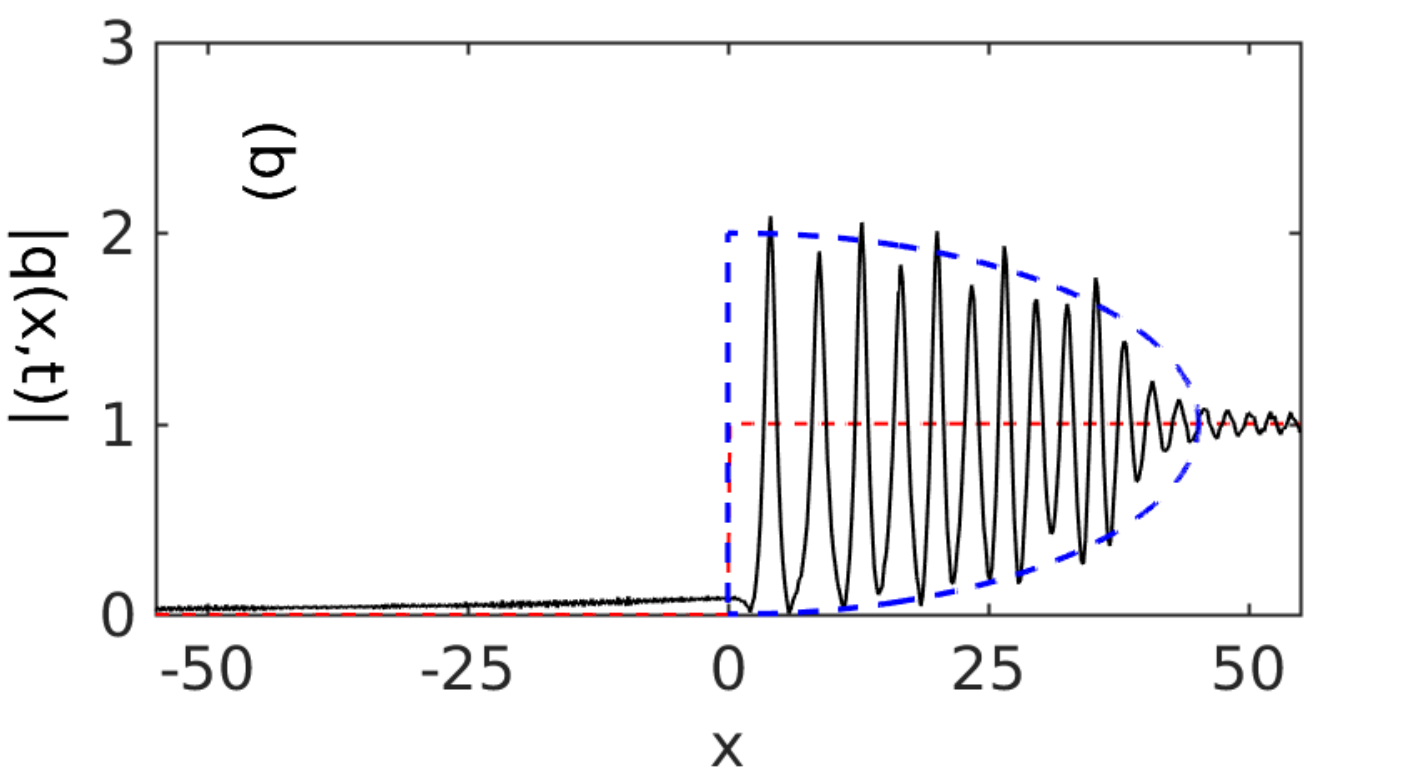}
\caption{%
(a) Density plot of the amplitude $|q(x,t)|$ of the solution of the focusing NLS Eq.~\eqref{e:nls} with $\epsilon=1$
generated by a one-sided step IC~\eqref{e:IC} with $A_-=\mu=0$ and $A_+=1$,
together with the boundaries (red lines) between the genus-0 and genus-1 regions predicted by Whitham theory.
(b) $|q(x,t)|$ (solid black line) as a function of $x$ at $t=8$,
together with the IC~\eqref{e:IC} (dashed red line) and
the envelope of the oscillations predicted by Whitham theory (dashed blue lines).
}
\label{f:1s}
\vskip1.6\bigskipamount
\centerline{\includegraphics[width=\figwdlt]{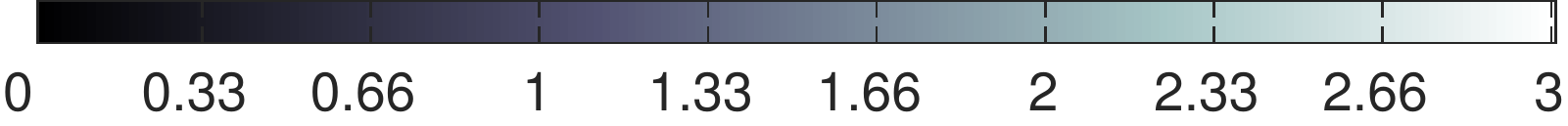}}
\caption{The grayscale used for all the density plots in this work.}
\label{f:colormap}
\kern-\smallskipamount
\end{figure}

The actual behavior of the solutions of the NLS equation with the above IC 
\cite{numerics}
is shown in Fig.~\ref{f:1s}
together with the predictions from Whitham theory,
demonstrating excellent agreement.
Importantly, note that the velocity of the matching point between the two solutions is zero, 
i.e., the discontinuity is pinned at $x=0$,
unlike what happens in the defocusing case \cite{OL20p2291}.

%%%%%%%%%%%%%%%%%%%%%%%%%%%%%%%%%%%%%%%%%%%%%%%%%%%%%%%%%%%%%%%%%%%%%%%%%%%%%%%%%%%%%%%%%%%%%
%%%%%%%%%%%%%%%%%%%%%%%%%%%%%%%%%%%%%%%%%%%%%%%%%%%%%%%%%%%%%%%%%%%%%%%%%%%%%%%%%%%%%%%%%%%%%

\paragraph{Symmetries}
Importantly, 
the invariances of the NLS equation induce corresponding symmetries for the Whitham equations.
%We already mentioned the invariance under phase transformations of the IC.
%Moreover, the solution~\eqref{e:modulationsystem} is easily generalized using the invariance of the NLS equation
%under scaling, spatial reflections and Galilean boosts.
%
(This is similar to what happens for other nonlinear evolution equations, even in two spatial dimensions, 
e.g., see \cite{PRSA2017}.)

Specifically, the Whitham modulation equations (and therefore the Riemann invariants) 
are insensitive to (i.e., invariant under) uniform phase rotations of the solutions of the NLS equation
[i.e., under transformations $q'(x,t) = q(x,t)\,\e^{i\phi}$, with $\phi$ an arbitrary real constant].
Spatial translations of the solution of the NLS equation 
[i.e., transformations $q'(x,t) = q(x-X_o,t)$, with $X_o$ an arbitrary real constant]
simply result corresponding translations of the Riemann invariants
[that is, $\alpha'(x,t) = \alpha(x-X_o,t)$ and $\gamma'(x,t) = \gamma(x-X_o,t)$].
Spatial reflections 
[i.e., the transformation $q'(x,t) = q(-x,t)$] 
simply yield a reflection of the  Riemann invariants
[that is, $\alpha'(x,t) = \alpha(-x,t)$ and $\gamma'(x,t) = \gamma(-x,t)$].
The invariance under scaling transformations
[i.e., transformations $q'(x,t) = a q(ax,a^2t)$, with $a$ an arbitrary real constant]  
yields a corresponding scaling for the Riemann invariants
[that is, $\alpha'(x,t) = a\alpha(ax,a^2t)$ and $\gamma'(x,t) = a\gamma(ax,a^2t)$].
Finally, 
Galilean boosts also translate to a corresponding transformation for the Riemann invariants.
That is, letting $q'(x,t) = q(x-2V_ot,t)\,\e^{i(V_ox-2V_o^2t)}$ 
(with $V_o$ an arbitrary real constant)
yields 
$\gamma'(x,t) = V_o + \gamma(x-2V_o,t)$ and 
$\alpha'(x,t) = V_o + \gamma(x-2V_o,t)$.
(Interestingly, the induced transformation for the Riemann invariants is essentially the same as in the corresponding symmetries 
of the Whitham equations for the KdV equation,
even though the transformation of the solutions of the two PDEs is very different \cite{EH2016}.)

The combination of the above invariances and the pinning of the discontinuity at $x=0$ 
for the self-similar solution~\eqref{e:modulationsystem} arising from the one-sided step IC~\eqref{e:IC}
is the key 
that enables us to use the above solution to analyze the more complicated scenarios,
as we discuss next.
In particular, note that 
multiplying Eq.~\eqref{e:IC} by $\e^{iV_ox}$ results in a slanted oscillatory wedge with 
boundaries at $x = V_\pm t$, where now $V_-= 2V_o$ and $V_+ = 2V_o + 4\sqrt2 A_+$.

%%%%%%%%%%%%%%%%%%%%%%%%%%%%%%%%%%%%%%%%%%%%%%%%%%%%%%%%%%%%%%%%%%%%%%%%%%%%%%%%%%%%%%%%%%%%%
\paragraph{Symmetric two-sided step}
Consider now a two-sided step given by the IC~\eqref{e:IC} with 
$A_- = A_+$ and $\mu = 0$,
corresponding to an initial phase discontinuity at $x=0$.
Without loss of generality we can set $A_\pm=1$
thanks to the scaling invariance of the NLS equation.

One can view the above IC as a superposition of two one-sided steps.
Of course the solution of the NLS equation is not simply given by a superposition of the corresponding one-sided solutions
in general.
Nonetheless, the property holds for the Whitham system in this case, thanks to the invariances of the one-sided solution,
the fact that the discontinuity of one-sided wedge is pinned at the origin,
and that the solution to the left of the wedge is asymptotically zero.
Thus,
the solution of the genus-1 Whitham system generated by the above IC 
is simply the linear superposition of the evolution of the one-sided step discussed above
and that of a reflected step.

\begin{figure}[t!]
\leftrightfigurepanel{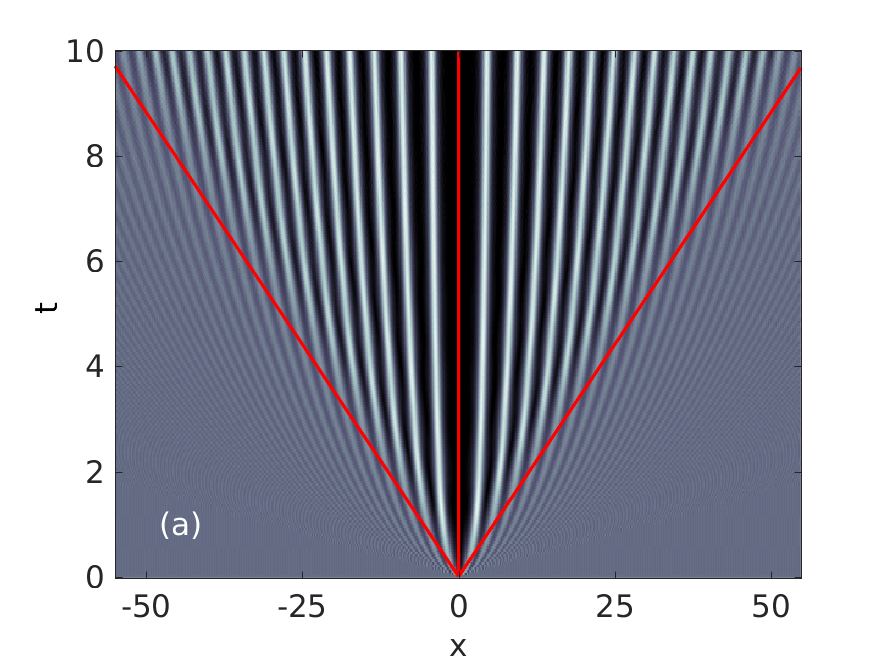}{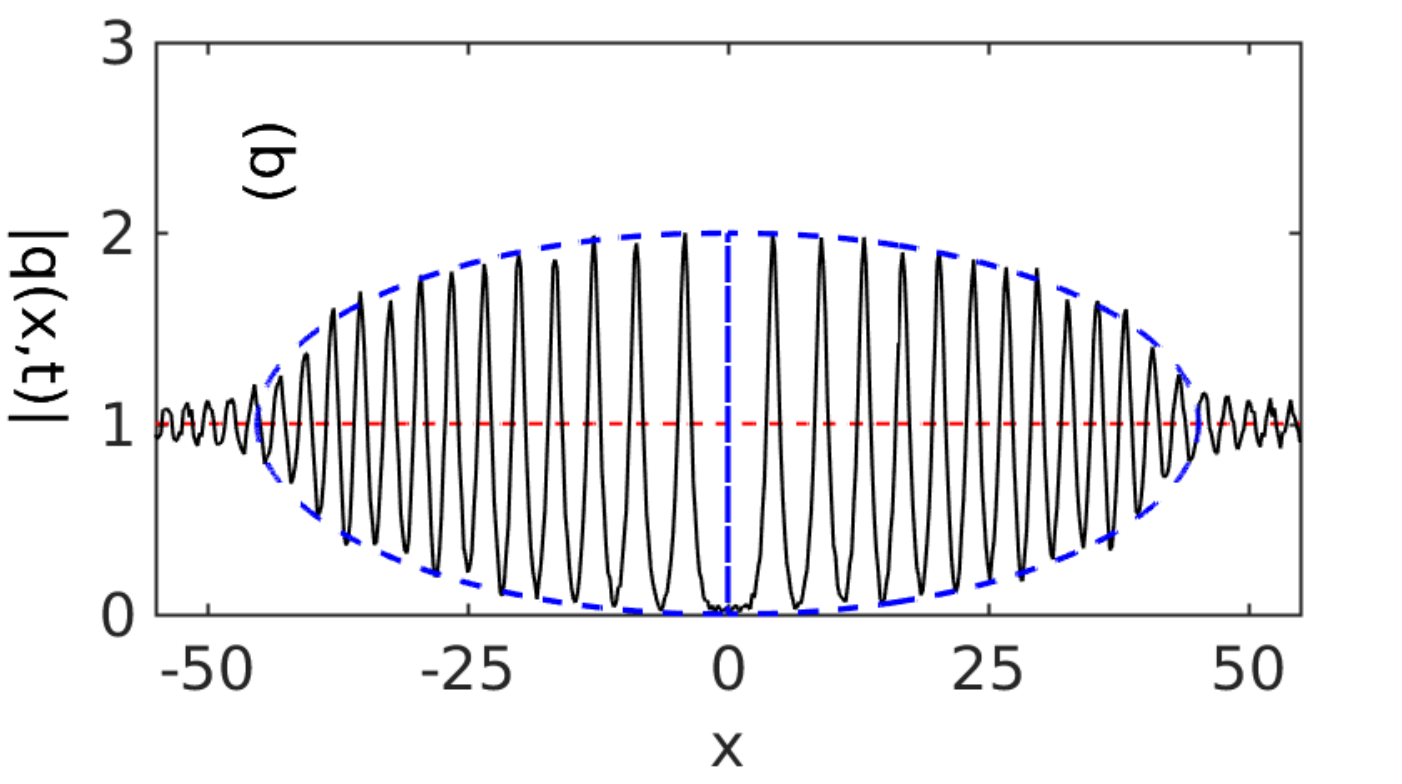}
\caption{%
Same as Fig.~\ref{f:1s}, but for a two-sided step IC given by~\eqref{e:IC}
with $A_- = A_+ = 1$, $\phi = \pi/2$ and $\mu=0$.
}
\label{f:2s}
\bigskip
\leftrightfigurepanel{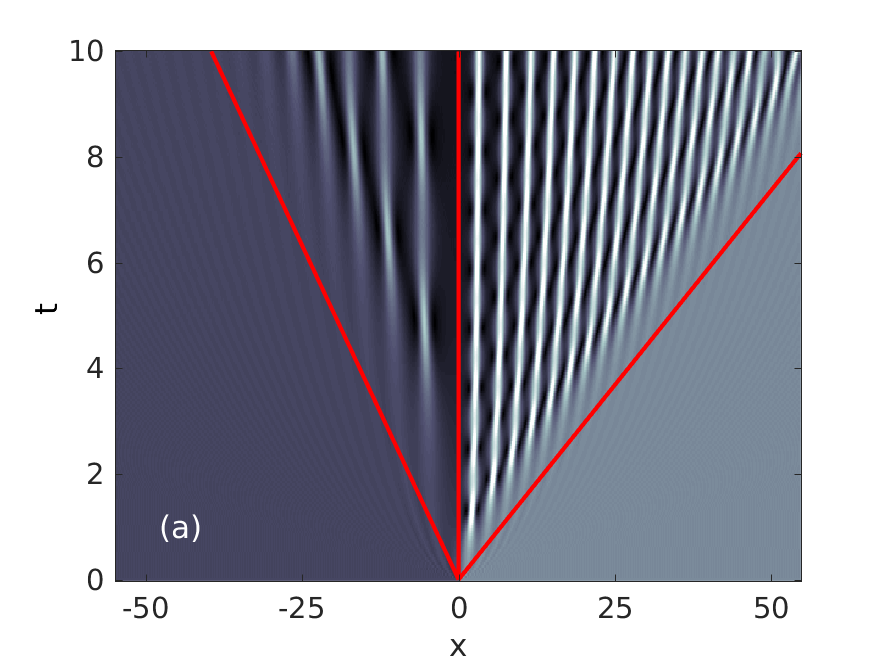}{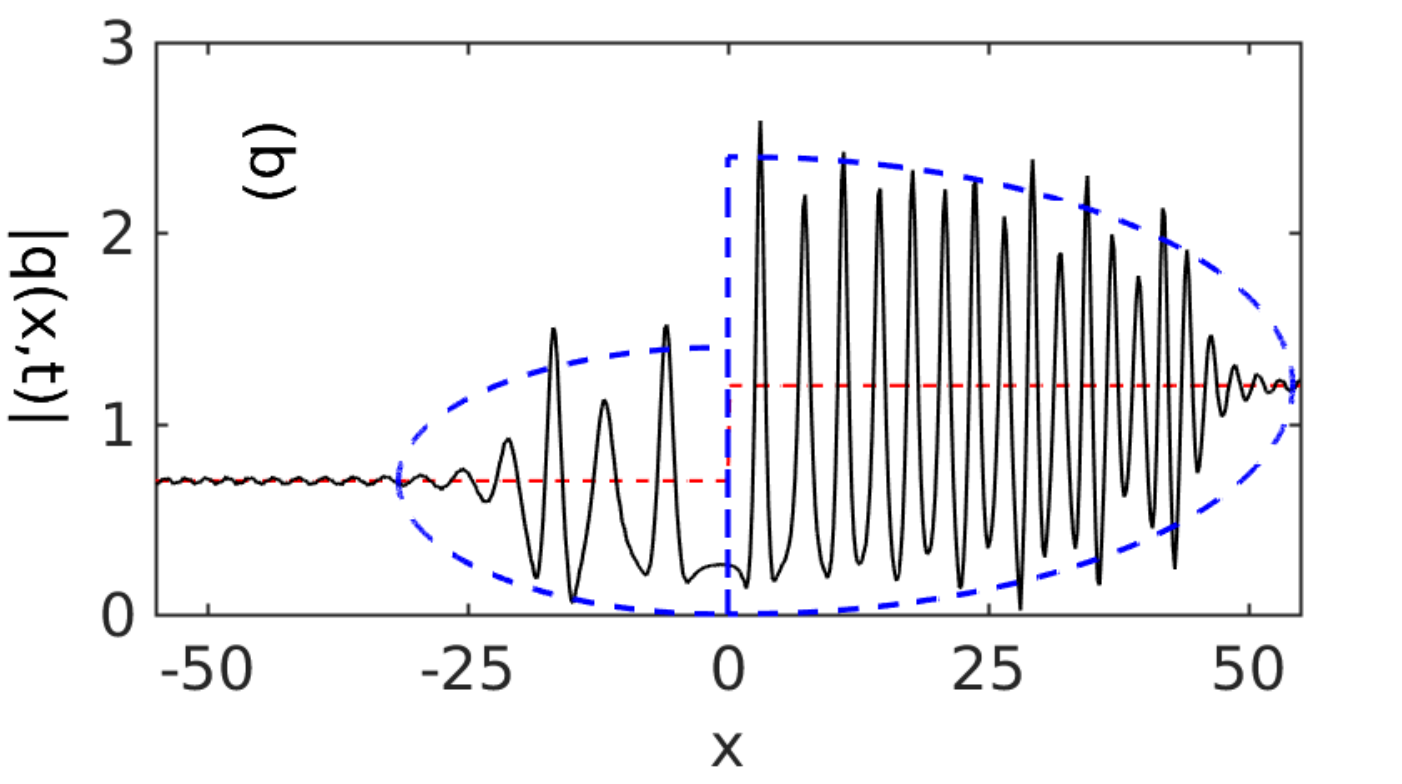}
\caption{%
Same as Fig.~\ref{f:1s}, but for an asymmetric two-sided step IC~\eqref{e:IC} with $A_- = 0.7$, $A_+ = 1.2$ and $\mu=\phi = 0$.
}
\label{f:asymm}
%\kern-4\medskipamount
\end{figure}

The resulting behavior is shown in Fig.~\ref{f:2s}, 
again demonstrating excellent agreement. %with the predictions of Whitham theory.
Note that the Whitham system is insensitive to the phase of $q(x,t)$ and hence to the value of $\phi$.
In particular, when $\phi=0$ the IC has no discontinuity at $x=0$; this is the case considered in \cite{elgurevich}.
But a similar behavior arises in the presence of a jump discontinuity in the phase of the IC.

It was suggested in \cite{elgurevich} that the self-similar solution~\eqref{e:modulationsystem} describes the evolution of
small perturbations of the constant background as a result of modulational instability.
And indeed it was shown in \cite{BM2016,BM2017} using IST that the long-time asymptotics of a very broad class of IC corresponding to localized perturbations
of a constant background tends asymptotically to this behavior as $t\to\infty$.
The properties of the self-similar solution were further studied in \cite{PRE94p060201R}.
Next we show how suitable combinations of the above one-sided solutions allow one to study even more general scenarios.

%%%%%%%%%%%%%%%%%%%%%%%%%%%%%%%%%%%%%%%%%%%%%%%%%%%%%%%%%%%%%%%%%%%%%%%%%%%%%%%%%%%%%%%%%%%%%
%%%%%%%%%%%%%%%%%%%%%%%%%%%%%%%%%%%%%%%%%%%%%%%%%%%%%%%%%%%%%%%%%%%%%%%%%%%%%%%%%%%%%%%%%%%%%

%%%%%%%%%%%%%%%%%%%%%%%%%%%%%%%%%%%%%%%%%%%%%%%%%%%%%%%%%%%%%%%%%%%%%%%%%%%%%%%%%%%%%%%%%%%%%
\paragraph{Asymmetric two-sided step}
We now consider the case $A_\pm\ne0$ and $A_-\ne A_+$, which we refer to as an asymmetric step.
Without loss of generality we can take $A_-<A_+$,
owing to the invariance of the NLS equation with respect to spatial reflections.

As before, one can view these IC as a superposition of 
two one-sided steps, but now with different amplitude.
As a result, the solution of the corresponding IVP for the Whitham equation is again given by 
the superposition of two one-sided self-similar solutions.
Note however that now the two halves have a different amplitude and different propagation speed.
An example of the resulting behavior is shown in Fig.~\ref{f:asymm} for $A_- = 0.7$ and $A_+ = 1.2$, 
demonstrating once more excellent agreement between the self-similar solutions of the Whitham equations 
and the numerical solution of the NLS equation.

%%%%%%%%%%%%%%%%%%%%%%%%%%%%%%%%%%%%%%%%%%%%%%%%%%%%%%%%%%%%%%%%%%%%%%%%%%%%%%%%%%%%%%%%%%%%%
\begin{figure}[b!]
\leftrightfigurepanel{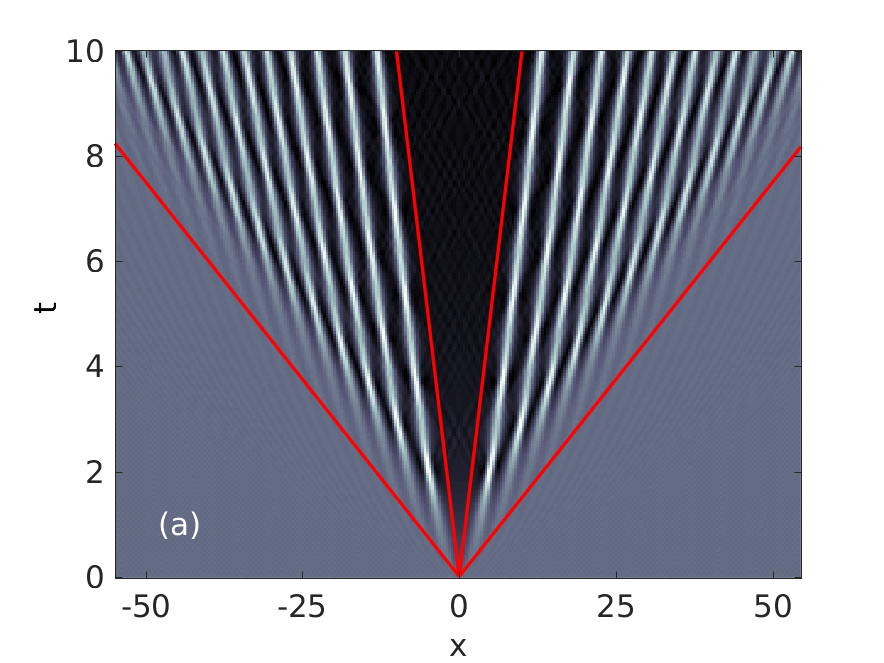}{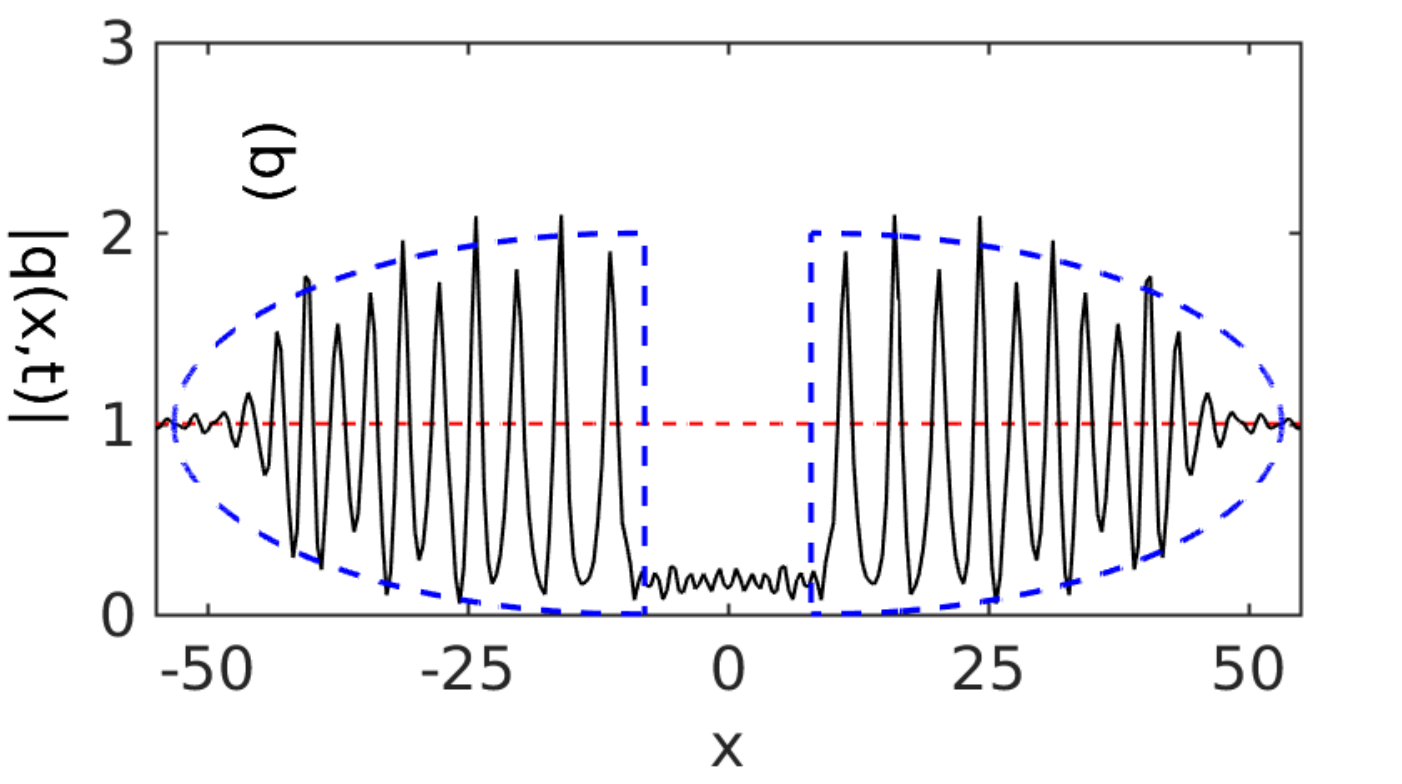}
\caption{%
Same as Fig.~\ref{f:1s}, but for outward counter-propagating flows given by the IC~\eqref{e:IC} with $A_\pm=1$ and $\mu=-0.5$.
}
\label{f:outflow}
\bigskip
\leftrightfigurepanel{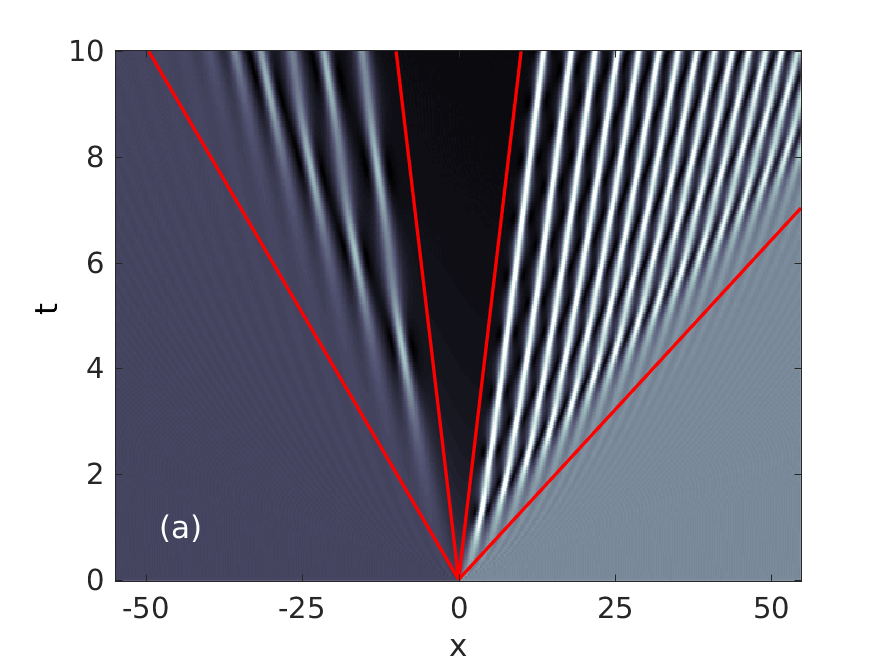}{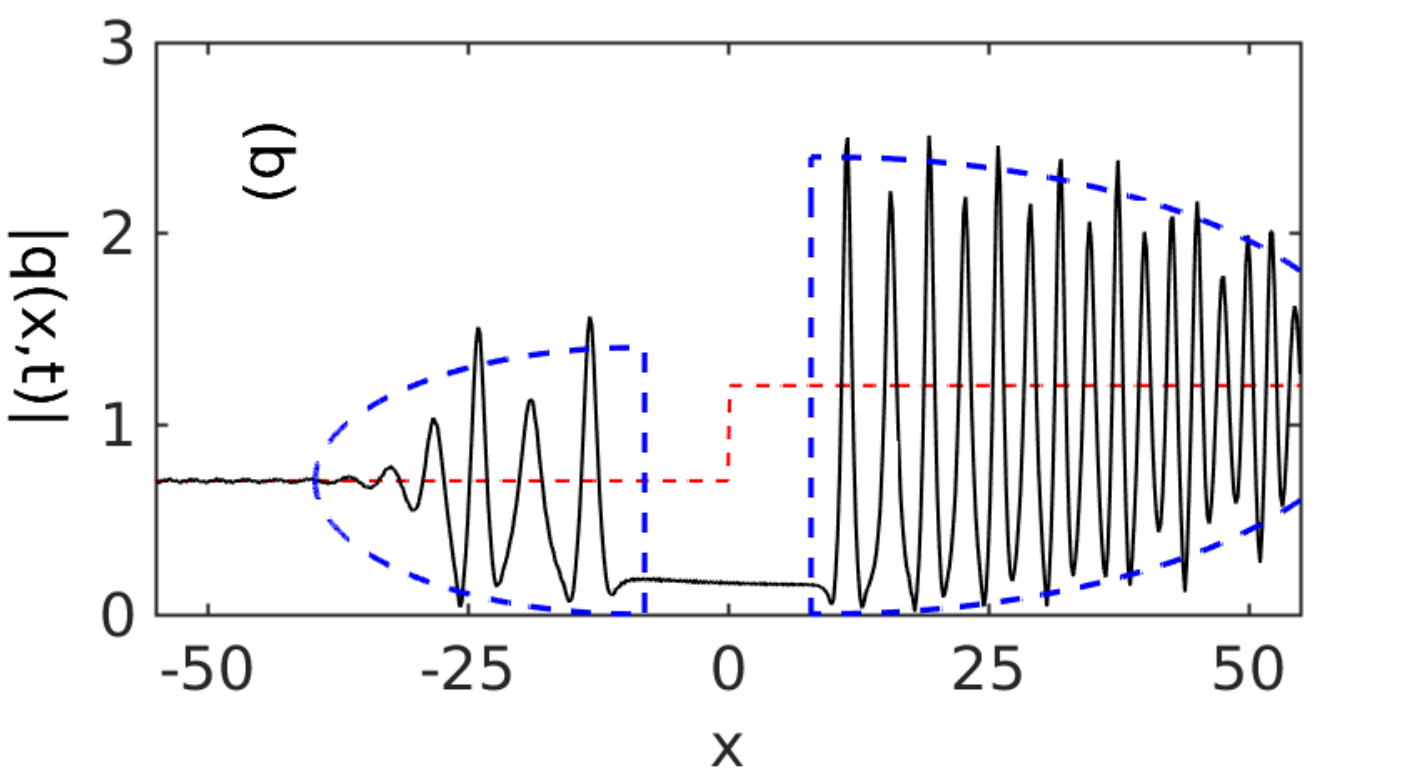}
\caption{%
Same as Fig.~\ref{f:1s}, but for an IC consisting of outward counter-propagating flows with unequal amplitude, 
given by Eq.~\eqref{e:IC} with $A_- = 0.7$, $A_+ = 1.2$ and $\mu=-0.5$.
}
\label{f:asymoutflow}
\end{figure}
%%%%%%%%%%%%%%%%%%%%%%%%%%%%%%%%%%%%%%%%%%%%%%%%%%%%%%%%%%%%%%%%%%%%%%%%%%%%%%%%%%%%%%%%%%%%%

%%%%%%%%%%%%%%%%%%%%%%%%%%%%%%%%%%%%%%%%%%%%%%%%%%%%%%%%%%%%%%%%%%%%%%%%%%%%%%%%%%%%%%%%%%%%%
\paragraph{Outward counter-propagating flows}
We now consider a different generalization of a symmetric two-sided step by allowing for the presence of counter-propagating flows, 
which are obtained when $\mu\ne0$.  
We first discuss the case of symmetric outward flows;
that is, we consider the IC~\eqref{e:IC} with $A_-= A_+ = A$ and 
$\mu<0$.
Again, without loss of generality we can take $A=1$ thanks to the scaling invariance of the NLS equation.

As before, it is useful to look at the IC as a superposition of two one-sided steps.
Because of the presence of a non-zero wavenumber, however, in this case the discontinuity for each solution half does not remain pinned at $x=0$,
but instead travels to the left or to the right with speed $2|\mu|$.
More precisely, the discontinuity in the left half is now located at $x= -2|\mu|t$ and the one in the right half at $x = 2|\mu|t$.
As a result, a wedge-shaped vacuum zone develops in the central portion of the $xt$-plane, i.e., for $|x|<2|\mu|t$.
An example of the behavior resulting from the above IC is shown in Fig.~\ref{f:outflow} for $\mu = -0.5$, 
demonstrating again excellent agreement with the corresponding solution of the Whitham equations.

A similar outcome is obtained when the IC combine outward counter-propagating flows
with an asymmetric step (i.e., an amplitude discontinuity), 
i.e., from Eq.~\eqref{e:IC} with $0<|A_-|<|A_+|$ and $\mu < 0$.
Indeed, Fig.~\ref{f:asymoutflow} 
shows such a case, obtained with $A_- = 0.7$, $A_+ = 1.2$ and $\mu = -0.5$,
demonstrating once more excellent agreement between the Whitham equations 
and numerical solutions of the NLS equation.

%%%%%%%%%%%%%%%%%%%%%%%%%%%%%%%%%%%%%%%%%%%%%%%%%%%%%%%%%%%%%%%%%%%%%%%%%%%%%%%%%%%%%%%%%%%%%
\begin{figure}[t!]
\leftrightfigurepanel{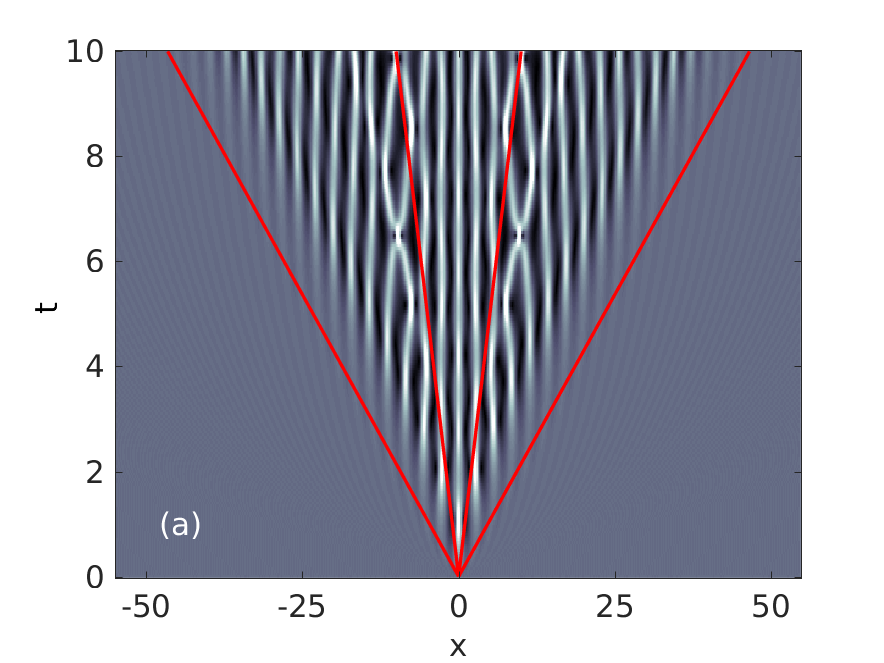}{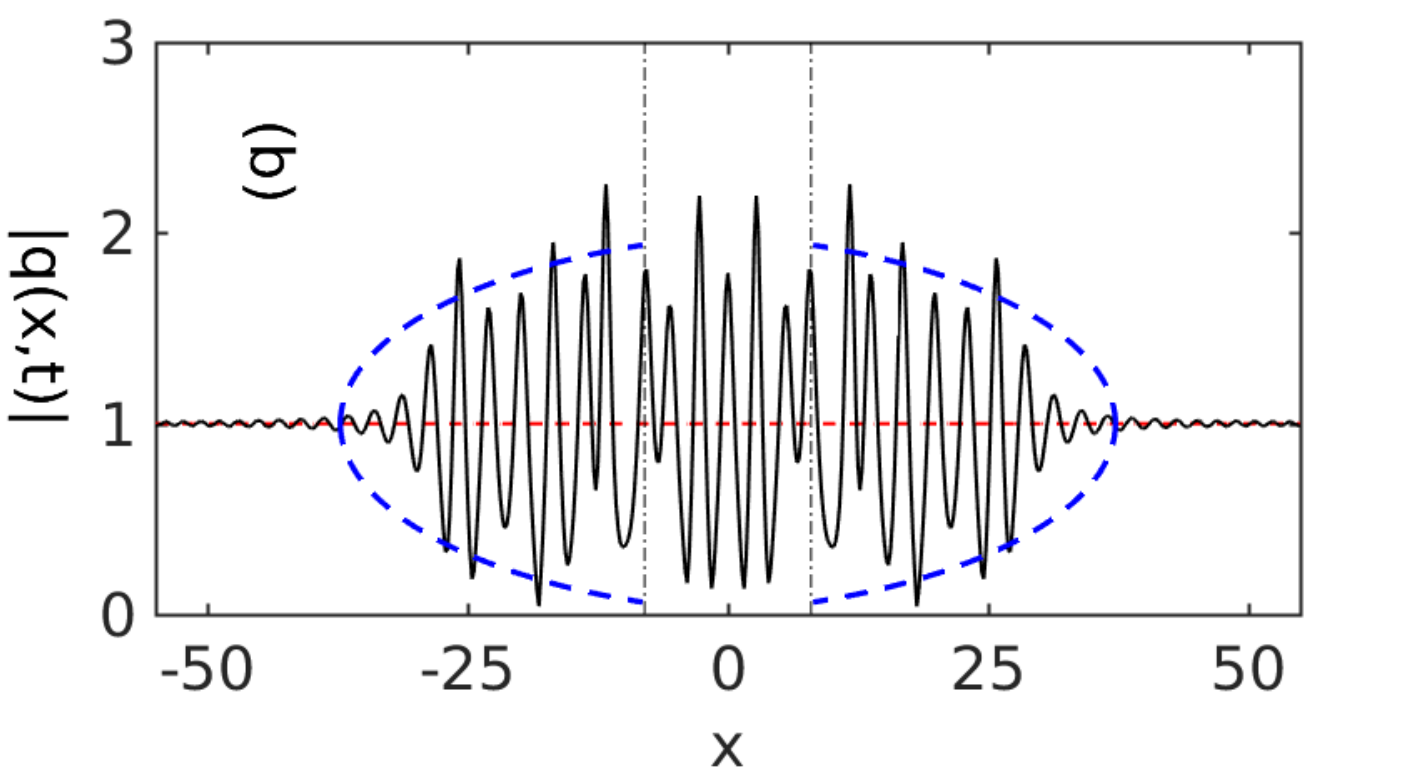}
\caption{%
Same as Fig.~\ref{f:1s}, but for inward counter-propagating flows given by the IC~\eqref{e:IC} with $A_\pm=1$ and $\mu=0.5$.
}
\label{f:inflow}
\bigskip
\leftrightfigurepanel{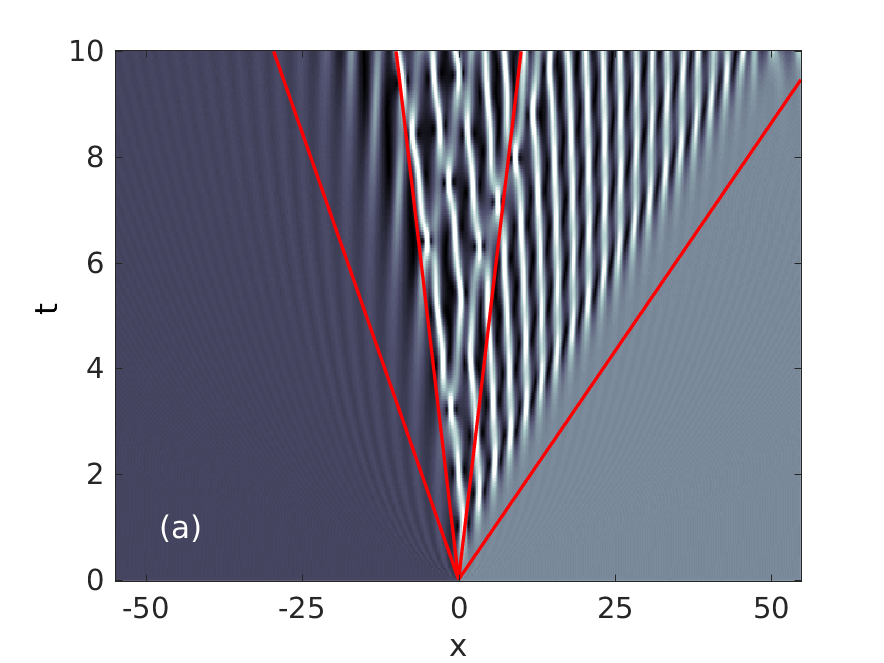}{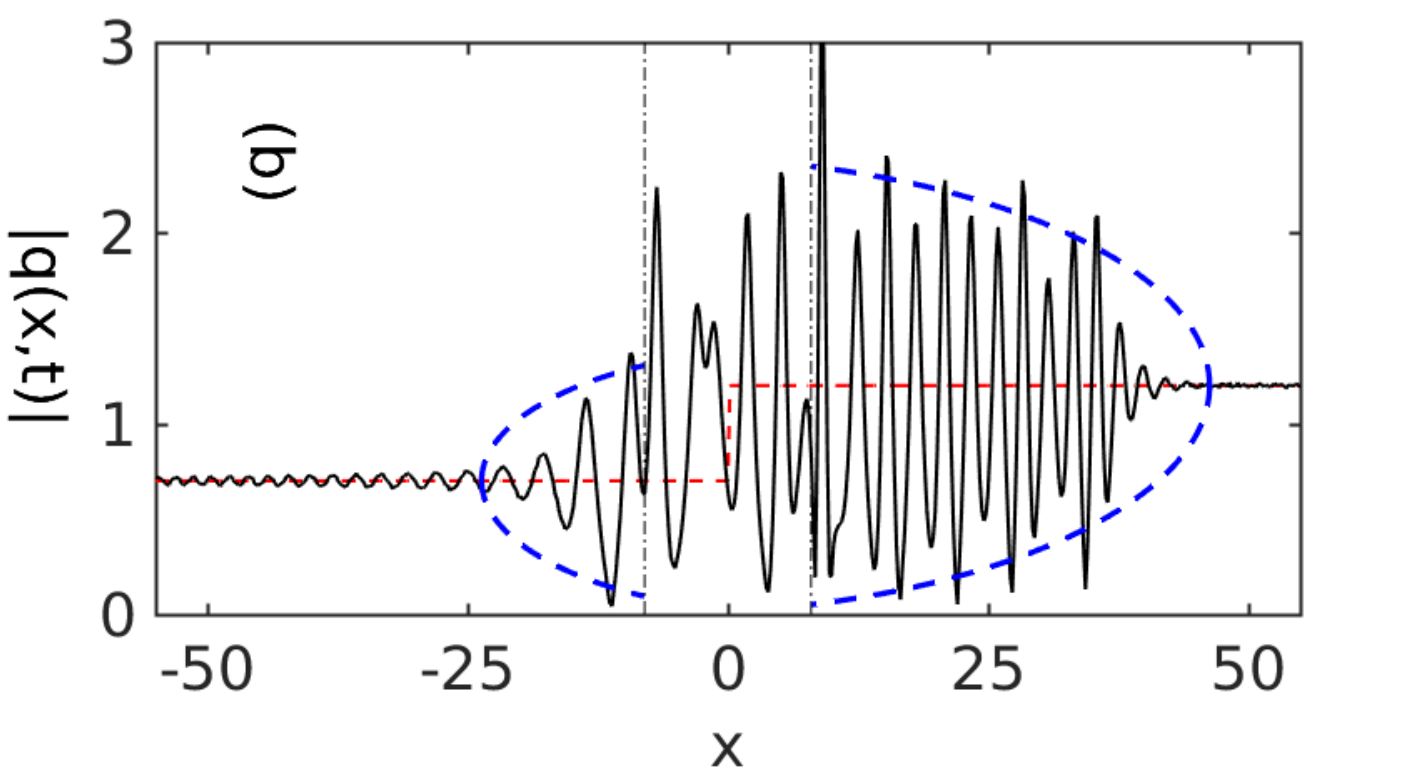}
\caption{%
Same as Fig.~\ref{f:1s}, but for an IC consisting of inward counter-propagating flows with unequal amplitude, 
given by Eq.~\eqref{e:IC} with $A_- = 0.7$, $A_+ = 1.2$ and $\mu=0.5$.
}
\label{f:asyminflow}
%\kern-\medskipamount
\end{figure}
%%%%%%%%%%%%%%%%%%%%%%%%%%%%%%%%%%%%%%%%%%%%%%%%%%%%%%%%%%%%%%%%%%%%%%%%%%%%%%%%%%%%%%%%%%%%%

%%%%%%%%%%%%%%%%%%%%%%%%%%%%%%%%%%%%%%%%%%%%%%%%%%%%%%%%%%%%%%%%%%%%%%%%%%%%%%%%%%%%%%%%%%%%%
\paragraph{Inward counter-propagating flows}
A different outcome is obtained when inward counter-propagating flows are present, namely when $\mu>0$.
%If needed, we can again set $A=1$ without loss of generality.
In this case, the non-zero carrier causes the two halves of the solution travel toward each other. 
In particular, the two individual solutions overlap in the region $|x|<2\mu t$.
%In this region, the small-dispersion limit is not described by solutions the genus-1 Whitham equations, 
%and is presumably described by a slow modulation of the genus-2 solutions of the NLS equation.

An example of the resulting behavior for $A_\pm =1$ and $\mu = 0.5$ is shown in Fig.~\ref{f:inflow}.
The dot-dashed lines in Fig.~\ref{f:inflow}(b) indicate the boundary of the overlap region. 
Inside this region, the interaction between the two genus-1 regions presumably results in the formation of a genus-2 region
(similarly to what happens in the defocusing case \cite{BK2006,HA2007}),
so the asymptotic expression for the solution in this region will be described by a slow modulation of the genus-2 solutions of the NLS equation,
and one cannot obtain useful information in this region using the genus-1 Whitham equations.

%Once again, 
%the predictions from Whitham theory are shown in Fig.~\ref{f:inflow} and compared with those from the numerical solution of the NLS equation.
Regions with more complicated oscillation patterns are indeed clearly visible in Fig.~\ref{f:inflow}.
(The appearance of genus-2 regions for this case had been predicted in \cite{BV2007}
based on the calculation of the long-time asymptotics of solutions.
Note, however, that \cite{BV2007} predicted a central genus-1 region surrounded by two genus-2 regions, 
each of which directly adjacent to the outermost genus-0 regions,
in constrast with the predictions from Whitham theory and the numerical results.
Moreover, the boundary between the two genus-2 regions and the genus-0 regions was predicted to be at $x = \pm 2(\mu + A^2/\mu)\,t$,
which is inconsistent with the limit $\mu\to0$, since in that case the solution reduces to the symmetric two-sided step 
discussed earlier.)
We also note that an even more complex scenario is obtained when $\mu>2\sqrt2 A$,
since in that case the genus-1 Whitham equations predict that the DSW region generated by the left half of the IC 
would be completely to the right of the one generated by the right half, and vice-versa.
It is not possible to make any predictions for this case using only the genus-1 Whitham equations.

The last scenario we consider is that of an asymmetric step with inward counter-propagating flows, 
which is obtained from Eq.~\eqref{e:IC} by taking $0<A_-< A_+$ and $\mu > 0$.
This case yields a similar outcome as that of a symmetric step, the only differences being 
once again the amplitude inside the oscillation region and its expansion speed.
The corresponding solution is shown Fig.~\ref{f:asyminflow}.
Interestingly, however, the difference between the genus-1 and genus-2 regions 
is more clearly identifiable in this case compared to that of a symmetric step.
We do not have a simple explanation for why this should be the case.
As before, more complicated outcomes can be obtained when larger
values of $\mu$ are considered.
In this case, however, we expect two different thresholds, 
one at $\mu = 2\sqrt2 A_-$ and one at $\mu = 2\sqrt2 A_+$.

%%%%%%%%%%%%%%%%%%%%%%%%%%%%%%%%%%%%%%%%%%%%%%%%%%%%%%%%%%%%%%%%%%%%%%%%%%%%%%%%%%%%%%%%%%%%%
\paragraph{Discussion}
In summary, we proposed a classification of the dynamical scenarios generated by
a single jump in the IC for the focusing NLS equation in the semiclassical limit.
The first few cases had been studied before, either by Whitham theory or by IST. 
%and a few other cases were studied in \cite{bikbaev}.
In particular, the self-similar solution of the genus-1 Whitham equations derived in \cite{elgurevich} 
was used in \cite{bikbaev} to present a qualitative description 
of the solutions produced by the inward and outward counterpropagating flow ICs discussed above.
Note, however, that no actual solutions of the NLS equation (either exact or numerical) were reported in \cite{bikbaev}.
Thus, a quantitative comparison between the predictions of Whitham theory and the actual solutions of the NLS equation
was still missing.
This is surprising, since, in the defocusing case, similar problems, as well as much more complicated ones, 
have been well-characterized \cite{OL20p2291,SJAM59p2162,BK2006,HA2007}.

While in this work the problem was formulated in the framework of semiclassical limits, 
there is a well-known correspondence between small dispersion limits (described by Whitham theory) and long-time asymptotics, 
which applies whenever the IC considered are scale-invariant,  
and the IC~\eqref{e:IC} studied in this work do possess this property.

There are marked differences between the behavior resulting from an initial discontinuity in the focusing versus the defocusing case.
In the defocusing case, a single discontinuity generates at most genus-1 regions \cite{PHYSD1995v87p186},
and one can only obtain genus-2 regions when two or more such discontinuities are considered \cite{BK2006,HA2007}.
Here, instead, we have seen that expanding genus-2 regions are generated when the IC contains inward counter-propagating flows
(as in Figs.~\ref{f:inflow} and~\ref{f:asyminflow}). 

Unlike what happens in other situations in both the focusing and defocusing case
\cite{millerkamvissis1998,bronskikutz1999,KMM2003,BK2006,HA2007}, 
here the boundaries of the genus-2 regions are not curved, but straight lines instead.
This is because of the scaling invariance of the IC considered here, which makes the semiclassical limit equivalent to the long-time asymptotics.

Whitham theory appears to slightly overestimate the spatial extent of the oscillation regions in some cases.  
Recall that small deviations between the predictions of Whitham theory and the actual PDE behavior are known to arise in the linear limit \cite{EH2016} of the former.
Moreover, all numerical computations in this work were done with $\epsilon=1$, while 
Whitham theory is designed to capture the behavior of solutions as $\epsilon\to0$,
so one would not necessarily expect the latter to be effective for such large values of $\epsilon$.
Perhaps better agreement could be attained for smaller values of~$\epsilon$.  
But, given the ellipticity of the Whitham equations in the focusing case, 
we find it quite remarkable that Whitham theory is even effective at all in the various situations considered here. 

In fact, while in the defocusing case one can prove that the solutions of the Whitham equations provide an asymptotic approximation for 
the time evolution of the corresponding IC for the NLS equation, 
the same is not possible in the focusing case since in this case the Whitham equations are elliptic.
The situation is the same as in the cases studied in \cite{elgurevich,EKT2016}.
Therefore, a rigorous description of the problems studied here can be obtained by computing the long-time asymptotics via the IST,
which will also be necessary to quantitatively describe the solution in the various genus-2 regions as well as 
in the cases when the genus-1 Whitham equations yield no predictions whatsoever
(e.g., in the case of inward counter-propagating flows when $\mu > 2\sqrt 2 A$). 

On the other hand, we reiterate that since Whitham theory does not require integrability, the results of this work are expected to 
also be applicable to many other NLS-type evolution equations that are not integrable, such as the ones considered
in \cite{SIREV},
which means that they should be experimentally observable in one of the various physical 
settings in which these models arise.
Indeed, experimental observation of some of the scenarios discussed in this work have already recently been reported in~\cite{arxiv1805.05074}.
It is therefore hoped that the remaining ones will also be realized experimentally in the future.

%%%%%%%%%%%%%%%%%%%%%%%%%%%%%%%%%%%%%%%%%%%%%%%%%%%%%%%%%%%%%%%%%%%%%%%%%%%%%%%%%%%%%%%%%%%%%
\paragraph{Acknowledgment}
It is a pleasure to thank G.\ A.\  El, M.\ A.\ Hoefer, D.\ Mantzavinos and S.\ Trillo for many insightful conversations 
on related topics.
This work was partially supported by the National Science Foundation under grant number DMS-1614623.
%and DMS-1615524.

%%%%%%%%%%%%%%%%%%%%%%%%%%%%%%%%%%%%%%%%%%%%%%%%%%%%%%%%%%%%%%%%%%%%%%%%%%%%%%%%%%%%%%%%%%%%%

%~
%\newpage
%~\par\noindent
%\textbf{Agenda}
%\\[1ex]
%

% DONE:
%
% 20180427  Redraw the 1d figs s.t.\ the max of $|q(x,t)|$ is always 3.
% 20180427  Redraw the 1d figs s.t.\ the solution is plotted @ $t=8$ instead of $t=10$.
% 20180427  Add boundaries between genus-0, genus-1 and genus-2 regions in the 2d figures
% 20180427  Add envelope of oscillations from Whitham theory to the 1d figures
% 20180427  Recompute Fig.~\ref{f:asyminflow} with a larger spatial domain s.t.\ no junk appears at the edge.
% 20180427  Explain that all figs are generated with $\epsilon=1$.
% 20180430  Discuss all figs in the text.

% NO:
%
%Add average from Whitham theory to the 1d figures


\begin{thebibliography}{12}
\bibitem{numerics}
All numerical simulations of the NLS equation were performed using an eighth-order Fourier split-step code with 4096 spatial grid points,
spatial domain $[-200,200]$ and integration time step $10^{-4}$.
\end{thebibliography}
\end{document}